**Employing Bayesian Networks for the Diagnosis and Prognosis of Diseases: A Comprehensive Review**

Carlos Segundo Muñoz-Valencia[1], José Antonio Quesada[2,3], Domingo Orozco[2], Xavier Barber[1,3*]

1. Center of Operations Research. Miguel Hernandez University
2. Department of Clinical Medicine. Miguel Hernandez University
3. Joint Research Unit STAT-SALUT UMH-FISABIO. Miguel Hernandez University

* Corresponding author: xbarber@umh.es

## Abstract

**Background**: Bayesian Networks (BNs) are probabilistic graphical models that leverage Bayes' theorem to portray dependencies and cause-and-effect relationships between variables. These networks have gained prominence in the field of health sciences, particularly in diagnostic processes, by allowing the integration of medical knowledge into models and addressing uncertainty in a probabilistic manner.

**Objectives**: This review aims to provide an exhaustive overview of the current state of Bayesian Networks in disease diagnosis and prognosis. Additionally, it seeks to introduce readers to the fundamental methodology of BNs, emphasizing their versatility and applicability across varied medical domains.

**Methods**: Employing a meticulous search strategy with MeSH descriptors in diverse scientific databases, we identified 190 relevant references. These were subjected to a rigorous analysis, resulting in the retention of 60 papers for an in-depth review. The robustness of our approach minimizes the risk of selection bias.

**Results**: The selected studies encompass a wide range of medical areas, providing insights into the statistical methodology, implementation feasibility, and predictive accuracy of BNs, as evidenced by an average AUC exceeding 75%. The comprehensive analysis underscores the adaptability and efficacy of Bayesian Networks in diverse clinical scenarios.

**Discussion**: The majority of the examined studies demonstrate the potential of BNs as reliable adjuncts to clinical decision-making. The findings of this review affirm the role of Bayesian Networks as accessible and versatile Artificial Intelligence tools in healthcare. They offer a viable solution for tackling complex medical challenges, facilitating timely and informed decision-making under conditions of uncertainty.

**Conclusion**: The encompassing exploration of Bayesian Networks presented in this review highlights their significance and growing impact in the realm of disease diagnosis and prognosis. It underscores the need for further research and development to optimize their capabilities and broaden their applicability in addressing diverse and intricate healthcare challenges.

**INTRODUCTION**

In medicine, determining the cause of an event is often not straightforward. Experts find numerous sources of possible uncertainty, unclear and incomplete measurements, and events for which it is difficult to establish causal relationships. Accordingly, probabilistic graphs are useful tools to understand in a visual and intuitive way how each factor affects the rest of the system.

A Bayesian Network (BN) is a probabilistic graph that tries to reproduce a real system. Used in basic and applied research as well as in artificial intelligence, BNs are based on Bayes' theorem and visualized with a directed acyclic graph (DAG), which is used to represent dependence or cause-and-effect relationships between variables. In this way, new knowledge is generated under conditions of uncertainty, which makes it easier to decide and reason through probability theory [1] BNs were introduced in the 1980s as a formalism for representing and reasoning through problems involving uncertainty, by adopting probability theory as a basic framework [2]. Since then, they have not only been used successfully in medicine, they have also been central to the development of numerous applications in other fields, including environmental sciences [3], production processes [4], finance [5], the debugging of artificial intelligence programs [6], and genetics [7], among others.

In the health sciences, BNs are applied to improve treatments, diagnosis, and prognosis, helping physicians to make a reliable decision by enabling a faster and more accurate prediction [8]. According to [9], the reasoning behind the wide application of BNs in medical diagnosis resides in their ability to express expert knowledge as well as to model uncertainty and deal with incomplete data.

This literature review aimed to provide a comprehensive description of BNs and their applications in medicine in general and in diagnostic and prognostic processes in particular, as well as to show the advantages of this method compared to others from the fields of artificial intelligence or machine learning, areas which have been attracting more attention in recent times.

Prior to presenting the conducted searches and analyzing the content of the articles, we will provide a concise overview of Bayesian networks, their types, classifiers, and evaluation methods. The aim is to introduce the reader to the subject matter, thereby facilitating a better understanding of the results and terminology used in the articles that will be analyzed subsequently.

## Bayesian networks

We will now describe the basic concepts around BNs, the types of networks, and the special structures known as Bayesian classifiers. We also briefly address the topic of learning about both parametric and structural BNs, and finally we present the computation of a BN, describing software available for BN applications.

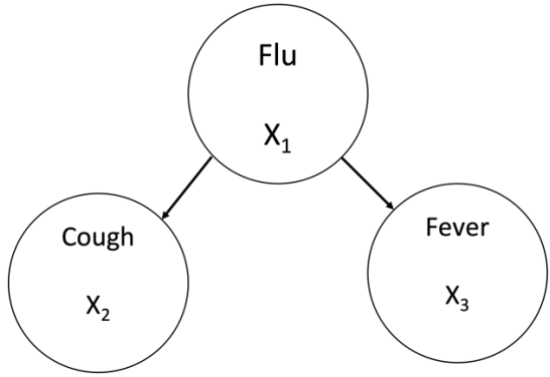

Figure 1. Representation of the relationships between variables (directed acyclic graph, DAG).

Let's start by looking at the example in Figure 1 where it defines a causal relationship between the variables flu, cough, and fever, where flu is expected to causally influence having a cough or fever. The variables cough and fever are not independent. If a person has a cough, they may have the flu, and then they probably have a fever.

However, consider that if a person has the flu, it is reasonable to conclude that the presence of fever does not depend on the presence of a cough. Therefore, we are going to assume that the variables cough ($X_2$) and fever ($X_3$) are conditionally independent given the variable flu ($X_1$). Formally, this can be expressed as follows [10]

$$P(X_2|X_1, X_3) = P(X_2|X_1)$$

$$P(X_3|X_1, X_2) = P(X_3|X_1)$$

therefore

$$P(X_1, X_2, X_3) = P(X_2|X_1)P(X_3|X_1).$$

The variable $X_1$ has an influence on both $X_2$ and $X_3$, but it is assumed that there is no direct relationship between $X_2$ and $X_3$. The representation of these relationships is usually done via a node and arrow diagram, connecting the influencing variables (primary variables) with the influenced variables (secondary variables). The structure shown in **Error! Reference source not found.** is a BN.

Considering the graphic structure of **Error! Reference source not found.** and more precisely the parents of each variable, the joint probability distribution can be written as follows:

$$P(X_1, X_2, X_3) = P(X_1|Pa(X_1)) \cdot P(X_2|Pa(X_2)) \cdot P(X_3|Pa(X_3))$$

where $Pa(X_i)$ is the parent variable of $X_i$.

This equation is the formal definition of a BN, in the case of three variables: using a process of analysis and classification of the unconditional independence between the three variables, we converted $P(X_1, X_2, X_3)$ into a product of three conditional probabilities.

Probabilistic networks are graphical representations of the variables and their relationships that characterize a situation [11]. Among the most commonly used are BNs [12] which provide information about the conditional relationships of dependence and independence present between the variables. The presence of independent relationships within a network makes BNs an excellent tool to represent knowledge in a consistent way, since the number of necessary parameters is reduced. These relationships abbreviate the representation of the joint probability function as the product of the conditional probability functions of each variable.

Because they represent a probability distribution, BNs confer a clear meaning, so they can be adapted for diagnostics, learning, explanation, and inferences [13]. Depending on the meaning given to them, they can represent causality [14] or correlation [15].

It could thus be said that a Bayesian network is a DAG whose nodes represent random variables and whose arcs represent direct dependencies, for example, causal relationships. Each node has a conditional probability table, which quantifies the relationship between the connected variables [16].

A BN is a probabilistic model made up of two different parts: on the one hand, there is the graphical structure (DAG), which defines the relationship between variables, and on the other hand, the probabilities established between these variables [17]. The elements of a BN are as follows [18]:

- A set of variables (continuous or discrete) that form the vertices or nodes of the network.
- A set of directed links connecting a pair of vertices. If there is a relationship with the direction $X \rightarrow Y$ then $X$ is said to be the parent of $Y$.

The network fulfills the following conditions:

- There is an association between each vertex $X_i$ with a conditional probability function $P(X_i|parents(X_i))$, which takes as inputs a specific set of values for the parent variables of the vertex and gives the probability of the variable representing $X_i$.
- The graph does not have directed cycles, that is, it does not have directed trajectories or paths that start or end at the same node.

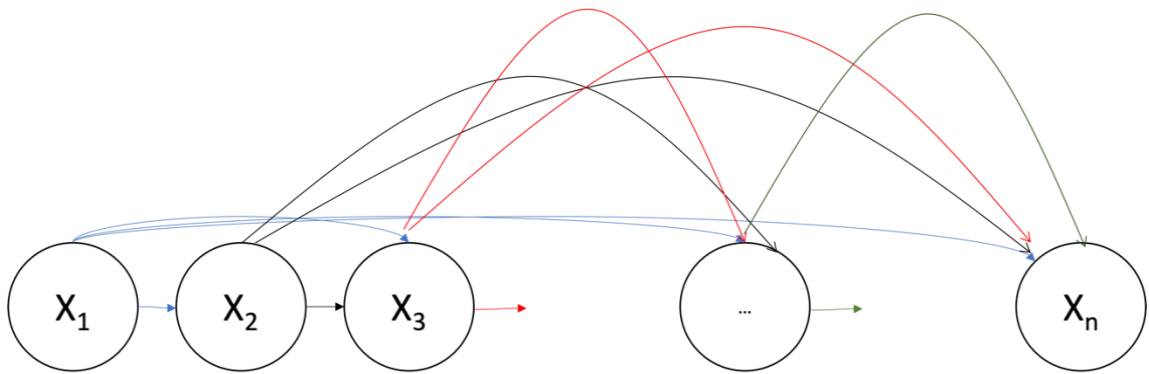

*Figure 2. Generic structure (DAG) of a BN for P with n random variables $X_1,...,X_n$.*

The BNs have a planning algorithm based on Bayes' theorem that describes and forecasts, so inferences can be made, calculating the posterior probability of unknown variables based on known variables. A BN provides an inference system that changes probability distributions as soon as new evidence is found about some vertices, while the new probabilities are transferred to the remaining nodes. The probability transfer is considered a probabilistic inference, which means that the probability of some variables can be determined from the evidence provided on other variables. Before the evidence is presented, they are called prior probabilities; afterwards, they are called posterior probabilities [18], [21].

Types of Bayesian networks

Depending on the type of random variables used in the construction of a BN, there are discrete BNs, where each variable takes a finite number of values and the probability distribution associated with each variable is multinomial. If the variables used to build the Bayesian networks are continuous, one method is to discretize the variables that is, divide them into nominal intervals. However, discretization implies a certain loss of information and the assignment of many parameters [22]. Modelling with continuous variables is a good alternative in this case. Gaussian BNs are used when continuous variables are used in the BN construction process, and each of them has a multivariate normal distribution [23]. When discrete and continuous variables are introduced in the construction process, they are called mixed BNs.

In discrete BNs, the probability distribution is determined by the probability tables, while in Gaussian BNs [24] it is determined by the joint density function.

For Dean et al. [25] in situations that evolve over time (for example, the sequential activation of brain areas during cognitive decision making), we need dynamic BNs. A previous BN specifies the initial conditions. In dynamic BNs, the time interval structures are identical, and the conditional probabilities are also identical over time. Therefore, dynamic BNs are time-invariant models, and dynamic just means that they can model dynamic systems.

Bayesian classifiers

Bayesian classifiers are a particular case of BNs, where there is a specific variable that is the class, and the other variables are attributes or characteristics. A Bayesian classifier determines the posterior probability of each class, $C_i$, using Bayes' theorem:

$$P(C_i|E) = \frac{P(C_i)P(E|C_i)}{P(E)}$$

Many Bayesian classifiers have been proposed for estimating probabilities [26], such as the naive Bayesian classifier (NBC), the tree augmented naive Bayesian classifier (TAN), and the Bayesian network augmented naïve Bayes (BAN) classifier.

The NBC allows the features to be independent of each other given the class, so the probability is found by multiplying the particular conditional probabilities of each feature given the class vertex [27]. $P(C)$ is the vector of prior probabilities of each class, and $P(E_i|C)$ is the conditional probability matrix of each feature. The features are conditionally independent given the class, so that there are no edges between them, as shown in **Error! Reference source not found.**.

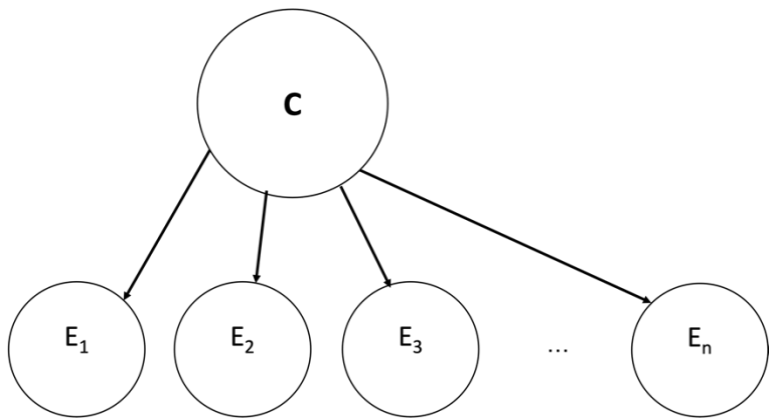

*Figure 3. An example of Naïve Bayesian classifier*

The NBC presents high classification precision in many areas [28]. However, sometimes its performance is affected by the dependence of the features or attributes. One way to solve this situation is to expand the basic NBC structure by adding edges between features. Two basic options are known [26]:

- The TAN, with a tree structure added between the characteristics, so that in principle there are few links and the complexity of the structure does not increase substantially.
- The BAN adds a general dependency structure between features, without restrictions.

### Learning Bayesian networks

Finding the structure that best fits the data becomes one of the problems to solve when using BNs. This task is performed by two types of machine learning: structural learning, which acquires the network structure (trees, polytrees, or multiconnected networks) from the data; and parametric learning, which has the associated probabilities (table of probability) of each root node and the other variables. In order to determine the probability distributions linked to each vertex of the network, it is necessary to know the type of network, that is, these two learning approaches cannot be carried out independently [27].

### Computation of a BN

The use of software plays an important role when building the structure of a BN, since it allows the graphic representation of a network so that it is understandable to those involved in its development. Madsen et al. [29] describe the software programs *Netica*, *Elvira*, and *Hugin*. Demo versions of *Netica* and *Hugin* are available for free, but some features are locked. Working with the demo version of the *Hugin* software allows a maximum of 20 variables, which is not enough to solve sequence analysis problems or even to compare results with other tools. Elvira is a software package for building and reasoning in probabilistic graphical models [30]. It has a somewhat academic look, speed, and reliability; moreover, it has incorporated several excellent ideas into its user interface. It allows an explanation of the model (static explanation) and an explanation of the inference (dynamic explanation). Elvira is the product of a joint project between several Spanish universities and is available for download.

The study by Charles River Analytics [12] explains the BNet program that applies BNs to predict and display weather forecasts. Murphy [31] provides different software applications on BNs. An a lot pages of internet present a private and free software.

Kenett [32] presents a list of software packages to generate and analyze BNs, including the following:

*GeNIe* is a development environment for reasoning in graphical probabilistic models, and SMILE is its inference engine. *GeNIe* is freely available for any use and has several useful features, such as a module for exploring and analyzing a data set and learning BNs and their numerical parameters from data. It has a special module that addresses problems related to diagnosis [10].

The *R* `bnlearn` package is powerful and free. Compared to other available Bayesian network software, it can perform more constraint-based and score-based methods.

*Bayesia* has also developed a proprietary technology for analyzing BNs. In collaboration with laboratories and large research projects, the company develops innovative technological solutions. Its products include 1) *BayesianLab*, a learning program 2) *Bayesia Market Simulator*, a simulation software package that can be used to compare the influence of a set of competing offerings relative to a defined population, 3) *Bayesian Engines*, a library of software components through which the modeling and use of Bayesian networks can be integrated, and 4) *Bayesian Graph Layout Engine*, a library of software components used to integrate automatic positioning of graphs into specific applications.

Model validation

The validation of the model justifies its correctness. Validation usually involves checking the performance of the model and its further improvement. Model performance can be expressed through several measures [10]as shown in Table 1.

|  | True Class (Gold Standard) | | |
|---|---|---|---|
|  | Positive | Negative | Measures |
| Predicted class — Positive |  | Positive | Negative |
| Predicted class — Negative | Positive | True Positive (TP) | False Positive (FP) |
| Predicted class — Measures | Negative | False Negative (FN) | True Negative (TN) |

*Table 1. Calculation of sensitivity, specificity, accuracy and positive and negative predicted value.*

*True positive (TP) – N items correctly recognized as positive; true negative (TN) – N items correctly recognized as negative; false positive (FP) – N items incorrectly recognized as positive; false negative (FN) – N items incorrectly recognized as negative; positive predictive value (PPV) – likelihood that a person who has a positive test result truly has the disease/condition; negative predictive value (NPV) – likelihood that a person who has a negative test result does not have the disease/condition.*

On some occasions, in addition to detecting the types of errors produced by the model, the test must meet high sensitivity and specificity requirements. In that regard, it is useful to take the measurements of the receiver-operating characteristics curve (ROC) [33], graphically observing the relationship between the two measurements. The area under the curve (AUC) provides a good measure of how well the BN model or classifier works. The closer to unity the area is, the more closely the model or classifier will conform to an ideal model or classifier (100% true positive and 0% false positive).

Given the information provided above, we trust that the reader will be able to comprehend all the articles that will be presented in the following sections.

### Bayesian networks in the health sciences

Since their creation, BNs have been linked to medical research, as probabilistic graphs are well suited to solving clinical problems. Understanding the causal relationships that lead to physiological processes is a challenge in clinical procedures; moreover, health sciences frequently require reasoning through difficult problems under conditions of uncertainty.

The probabilistic diagnosis has gained popularity in the medical literature in recent decades. According to Warner et al. [34], and Gorry and Bamett [35], Bayesian methods began to be adopted in diagnostic procedures in the USA and the UK in the 1960s. These diagnoses used frequentist probabilistic methods to select a variable that includes possible diagnoses, as well as several variables that typically correspond to symptoms. For their part, in the 1980s Schwartz et al., [36] and Spiegelhalter and Knill-Jones, [37] described the development of BNs and influence diagrams in the medical field, from their definition to the construction of algorithms for the efficient processing of certainty.

In applying probability graph models, there are three main reasons that a researcher encounters uncertainty and inaccuracy: randomness, lack of precision in data collection, and model flaws. In clinical procedures, this manifests in incomplete clinical histories, the subjective component presented by physicians or patients, and inaccuracies in measurements [38].

Two approaches are taken into account when building BNs: starting from a data set, through the application of techniques from network learning; or with the help of experts, where health professionals transfer their experience to include variables with cause-and-effect relationships in the model. The two approaches can be mixed by creating models using learning techniques, removing and imposing relationships based on personal criteria.

The first approach is the fastest way to build BNs in medicine, taking into account that the number of observations must be sufficiently large. However, it does have drawbacks: each observation often has only a few measurements in addition to the diagnosis. These variables are probably not enough to find their conditional independence, and they may not factor into the health professional's diagnosis.

In line with the above, there is usually an expert opinion on the model. The construction of BNs based on personal judgment can be divided into two stages: in the first, qualitative data are collected, clinically relevant variables are identified, and their relationships are used to build causal networks. In the second stage, quantitative information is collected: prior, conditional, and marginal probabilities.

Among specialists in statistics and information technology, interest has been growing in training doctors so they can appropriate the use of models. Accordingly, the BNs in medicine could face challenges, such as the construction of algorithms that allow them to reduce the degree of difficulty in the growth of the BNs [Arora et al, 2019].

Notably, probabilistic graph models are frequently used to forecast dependencies within a genetic test [7]. In parallel, Correa and Goodacre [39] have developed BNs that explain assimilation networks by the modification and action exerted reciprocally between metabolites. Similarly, in the area of neurology, dependency modeling in cellular tissues has been widely applied by means of BNs [22].

### Bayesian networks in the diagnosis and prognosis of diseases

Medical diagnosis is often simplified to a reasoning that entails the construction of hypotheses for each disease given the set of observed findings in a patient. The diagnosis results from choosing the most probable hypothesis for a set of observations. Formally, it can be expressed by the equation [10]:

$$\text{Diagnosis} = \max_i P(D_i|E)$$

$P(D_i|E)$ is the probability of the disease $D_i$ given the evidence $E$, which represents the set of observed findings, including signs and symptoms along with laboratory results.

For its part, medical prognosis tries to predict the patient's future state based on a set of observed findings and the administered treatment. Formally, it can be expressed by the equation:

$$\text{Prognosis} = P(O|E,T)$$

Variable $E$ is the evidence, that is, a set of observed findings such as signs, symptoms, and laboratory results; $T$ represents a prescribed treatment for a patient, and the variable $O$ is the outcome that can represent, for example, life expectancy, health status, or the spread of a disease.

Based on the information provided above, we are confident that the reader will be able to comprehend all the articles that will be presented in the following sections.

**MATERIAL AND METHODS**

The scope of this review is the recent fields of research on BNs and their applications in medicine, in general and more specifically in the diagnosis and prognosis of diseases.

### Data sources

The Web of Science, Scopus, ProQuest Central, Google Scholar, and PubMed Central were searched for published articles containing the following keywords: Bayesian networks, disease prediction, medical diagnosis, applications in medicine, prognosis of a disease. The search strategies implemented in this study were constructed utilizing predefined keywords and aligning them with the pertinent descriptors of the chosen databases. The searches covered literature from 1983 to the present to ensure an exhaustive inclusion of pertinent studies throughout this extensive timeframe. This methodology was designed to enable an in-depth review of the existing body of knowledge, thereby facilitating a nuanced comprehension of the advancements and discoveries in the field.

MeSH terms: Medical Informatics, Decision Making, Computes-Assited, Diagnosis, computer-assisted and Pronognis , computer-assisted.

### Data extraction

The search period was from 1983 to 2023. References of included articles were hand searched to identify additional relevant records based on the title, abstract and full text.

### Synthesis

Included studies were classified according to their main focus: (1) BNs, (2) application of BNs in health sciences, (3) and applications of BNs in the diagnosis and prognosis of diseases. Groupings were defined based on the data.

**RESULTS**

The searches yielded a total of 190 records, and following the screening process, 60 articles focusing on BNs applied to medicine for diagnosis and prognosis were finally included in the review (Figure 4)

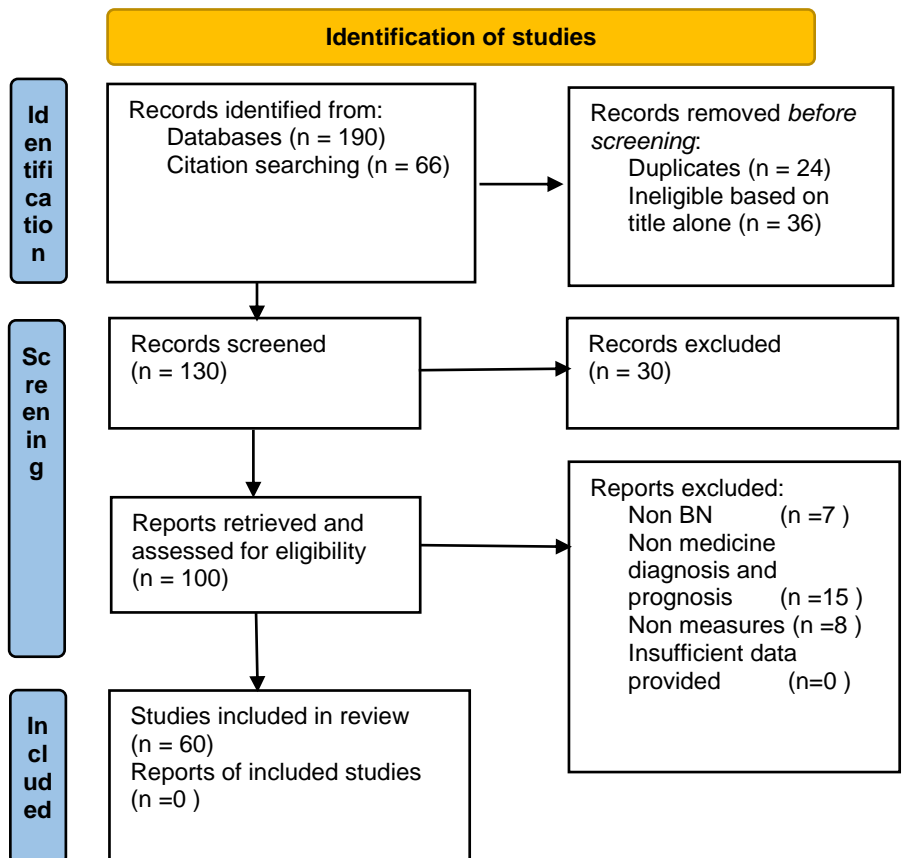

*Figure 4. PRISMA flow chart*

Many models of BNs have been developed for medical diagnosis. They can be built using data, knowledge, medical literature, ontologies, other sources of medical evidence, or a combination of these. For example, Moreira et al. [20]presented a BN model for the diagnosis of preeclampsia built on the basis of knowledge. Seixas et al. [[40] compared multiple models of BNs (developed from knowledge and data) with other classifiers fully learned from data to diagnose Alzheimer's, dementia, and mild cognitive impairment. For each of these conditions, the performance of the knowledge-based BNs models, with AUC values of 0.86, 0.96, and 0.97, respectively, was comparable to the best data-based classifier, with AUCs of 0.90, 0.98, and 0.97, respectively. In contrast, Yin et al. [41] analyzed the data to identify the most important variables and then trained a BN to diagnose Alzheimer's based on that information. Sa-Ngamuang et al. [42] presented a BN model to differentiate dengue from other acute febrile illnesses, combining expert knowledge, published evidence, and data. Farmer et al. [43] constructed a BN for diagnosing musculoskeletal disorders of the shoulder using expert knowledge and the medical literature, along with retrospective data. The expert elicitation was performed in two stages: an initial model was updated following a review by a panel of orthopedic specialists.

Some authors have suggested extracting medical knowledge from dictionaries or medical ontologies. Bucci et al. [44] built a two-level BN model for diagnosis that automatically extracted medical knowledge from an ontology; the diagnostic part was supplemented by a decision network to evaluate the available tests using utility values specified by a clinician. In Zhang et al. [45], several models, including a BN model, were automatically developed to diagnose mild cognitive impairment using a combination of approaches, including insights drawn from the SNOMED ontology.

Other studies have used only data to build BN models. For example, Moreira and Namen [20] combined structured data with information extracted from medical notes (i.e., text) to build a

BN model for diagnosing dementia. In another study, Somnay et al. [46] trained four data-based classifier models (rules, logistic regression, tree, and bay network) to recognize primary hyperparathyroidism.

## Detailed Examination of Key Studies

[47] presented an insight-based BN modeling approach to diagnose rheumatoid arthritis. The author explain that Bayesian network models have exhibited strong predictive abilities in forecasting osteoporosis in postmenopausal women. Moreover, these networks provide a more intuitive representation of the complex interplay of risk mechanisms between diseases and various factors. The authors say that Bayesian networks provide a more clear representation of the complex interaction of risk mechanisms between various factors and diseases. And other studiy with similar methodology studies bone mineral density using multiple linear regression and Bayesian network model [48].

Cruz-Ramirez et al. [49]evaluated the effectiveness of seven BN classifiers as potential tools for diagnosing breast cancer using two real-world databases containing fine needle aspirates of breast lesions, collected by a single and multiple observers, respectively. The results showed a certain element of subjectivity in the data: the single-observer data set showed a mean accuracy of 93.0%, compared to 83.3% for the multiple-observer data set. These findings suggest that observers see different things when examining samples under a microscope, a situation that significantly decreases the performance of these classifiers in the diagnosis of such a disease.

Lee et al. [50] proposed a BN model proposed a BN model to identify actionable factors contributing to racial differences in breast cancer stage at diagnosis. They "Employed diverse Bayesian model evaluation tools to assess these two hidden pathways, as well as each of the four observed variables, in elucidating racial disparities in the stage at diagnosis.

Yaneli et al [51]. used thermography for the prediagnosis of breast cancer based on the predictive value of its attributes. The BN used included a database of 98 patients. The results suggested that these attributes are not sufficient to produce good results in the prediagnosis of breast cancer. The models showed unexpected interactions between thermographic attributes, especially those directly related to the class variable. Djebbar and Merouani [52] used the BNs to model the case-based reasoning and applied it to the diagnosis of liver disease. They found that the BN was an excellent tool to model uncertainty in terms of its clear graphical representation and the laws of conditional probabilities based on the similarity function, and the BN was apt to select the most similar case as the recovery phase and optimize it. They used two exact inference algorithms, JLO and Pearl, to calculate conditional probabilities in order to improve the performance of case-based reasoning.

Vila-Francés et al. [17] presented a decision-making support plan, using a BN to predict the probability of having unstable angina based on clinical data. The final model was implemented as a web application that is currently being validated by clinical specialists. Seixas et al. [40] proposed a BN detection model for the early detection of Alzheimer's diseases, dementia, and mild cognitive impairment. The parameters of this structure, built by experts in this field, were estimated using learning algorithms from a real-world data set. The proposed BN showed better accuracy for diagnosing Alzheimer's, dementia, and mild cognitive impairment compared to most other known classifiers. In addition, it provided additional useful information to clinicians, such as the contribution of certain factors to the diagnosis. Zhou et al. [53] described the development of BiotinNet, an Internet-based program that uses BNs to integrate published data on various aspects of biotin metabolism. Users can provide a combination of values for biotin-related metabolite levels to obtain predictions for other metabolites that are not specified. This system can help researchers design future experiments while enabling the continuous incorporation of new data.

Shen et al. [54] explored the construction of a clinical BN for the probabilistic inference of medical ontology based on electronic medical records. By evaluating the learned topology against the opinions of expert clinicians and entropy calculations, and by calculating the diagnostic classification based on ontology, the study demonstrated that direct and automated construction of a high-quality health topology and ontology is feasible using medical records.

In recent years, many studies have been carried out on BNs for disease prognosis:

Cao et al. [55] compared BNs versus convolutional neural networks versus multivariate logistic regression for predicting long-term postoperative health status. The BN showed excellent predictive capacity for type 2 diabetes and dyslipidemia (AUC 0.94 and 0.92, respectively), good capacity for hypertension and sleep apnea syndrome (AUC 0.89 and 0.83, respectively), and fair capacity for depression (AUC 0.75). The study showed that BNs are useful tools to predict long-term health status and comorbidities in patients after bariatric surgery.

Siga et al. [56] proposed a BN model to predict all-cause mortality in hemodialysis patients. The model presented an AUC of 0.78 (standard deviation [SD] 0.01); a true positive rate (sensitivity) of 72% (SD 2%); a true negative rate (specificity) of 69% (SD 2%); a positive predictive value (PPV) of 70% (SD 1%); and a negative predictive value (NPV) of 71% (SD 2%). The BN provided the most reliable prediction when comparing the results with those acquired by logistic regression.

Spyroglou et al. [57], [58] examined the performance of BN classifiers for predicting asthma exacerbations based on various patient parameters, including objective measurements and medical history data. The proposed semi-naïve network classifier was able to predict whether a patient would have a disease exacerbation after the last evaluation with an accuracy of 93.8% and a sensitivity of 90.9%. In addition, the resulting structure and conditional probability tables gave a clear view of the probabilistic relationships between the factors. This network could help clinicians identify patients at high risk of exacerbation after stopping medication and confirm which factors are the most important.

Cai et al. [58] combined BNs with importance measures to identify key factors that have significant effects on the survival time of patients with hepatocellular carcinoma after hepatectomy. The accuracy of the BN model was 67.2%. The true positive rate (TPR) of the model was 83.22%, and the false positive rate (FPR) was 48.67%.

Verduijn et al. [59] presented a BN that implements a process-oriented and dynamic vision for forecasting. The procedure optimizes the prediction of results and accounts for the patient dropout at an early stage. The paper describes how prognostic BNs can be applied to solve a series of information problems related to medical prognosis.

Tian et al. [60] to Illustrate the multivariable probabilistic associations between CVD risk and metabolic health, alongside obesity status, and discern potential elements influencing these relationships among Chinese adults. The author conclude that utilizing network modeling is beneficial for amalgamating expert knowledge and observational data, facilitating the straightforward recognition of probabilistic dependencies and conditional independencies among variables via graphical representation.

Agrahari et al. [61] presented a method based on the analysis of BNs to classify two types of hematological malignancies: acute myeloid leukemia and myelodysplastic syndrome. The BN classifier showed an accuracy of 93% and a precision of 98%, outperforming the results achieved by other investigators on the same data set.

Elazmeh et al. [62] built a reliable BN model for early assessment of emergency pediatric asthma exacerbations. Their predictive model was capable of distinguishing between patients with mild or moderate/severe asthma attacks at a medically acceptable level of performance.

Wang et al. [63] proposed a BN model to describe and predict the occurrence of brain metastases from lung cancer, which outperformed the naïve Bayes, logistic regression, and support vector machine models in terms of average sampled sensitivity. Furthermore, the proposed BN had advantages over the other approaches for interpreting how brain metastases develop from lung cancer.

Wu et al. [64] developed a BN model to guide individually targeted antibiotic therapy at the point of care by predicting the most likely causative pathogen in children with osteomyelitis and the antibiotic with the optimal expected utility. BNs explicitly model the complex relationship between the unobserved infectious pathogen, observed culture results, and clinical and demographic variables, integrating data with critical expert knowledge under a framework of causal inference.

Takenaka et al. [65] used a BN to predict postoperative clinical recovery from foot drop attributable to lumbar degenerative diseases. The results obtained suggest that the clinician can intuitively understand the layered correlation between the predictors of the BN models, which successfully provide probability estimates of post-tibial anterior muscle strength to inform treatment.

Sesen et al.[66] assessed the feasibility of BNs for providing accurate and personalized survival estimates and treatment selection recommendations in lung cancer care using the Lung Cancer Database (LUCADA).

Prochazka et al. [67] presented BN modeling as a tool to predict early disease progression in patients with follicular lymphoma. This paper showed that the BN model has better prognostic power than multivariable logistic regression in terms of predicting disease within 24 months of first-line immunochemotherapy (POD24). Unlike the logistic regression model, BNs allow visualization of complex relationships between predictors and individualized prediction of the risk associated with POD24, even if some of the predictors are unknown.

Kaewprag et al. [68] developed a BN-based predictive model that enables clinicians to better understand and explore clinical patient data along with risk factors for pressure ulcers in intensive care unit patients based on electronic medical records. The BN model increases the sensitivity of the prediction compared to logistic regression models, without sacrificing overall accuracy.

Park et al. [69] built a prognostic model for metabolic syndrome using BNs with an evolutionary optimized ordering approach. This proposed model outperformed the conventional BN model and other neural network models.

Velikova et al. [70] presented a decomposition of the BN to model the detection of breast cancer, demonstrating some advantages of the method: natural and more intuitive representation of breast abnormalities and their characteristics; compact representation and efficient manipulation of large conditional probability tables; and a possible improvement in the processes of knowledge acquisition and representation.

Finally, Derevitskii et al. [71] developed a BN-based approach to predict important clinical indicators to improve treatment for COVID pneumonia. The models were validated using expert knowledge, current clinical recommendations, previous research, and classic predictive metrics. Following validation of other COVID pneumonia data sets for other hospitals, the proposed models can be used as part of decision-making support systems to improve treatment for COVID pneumonia.

## DISCUSSION

As observed in the articles cited in the preceding section, Bayesian Networks serve as a highly valuable tool within the realm of health sciences. They offer considerable assistance to medical professionals, boasting quite acceptable levels of accuracy. This section presents a discussion about the results obtained by BNs for disease diagnosis and prognosis, as well as relevant considerations, advantages, and disadvantages

Another important finding is the significant difference between COVID and non-COVID pneumonia. A more in-depth exploration into COVID studies has not been pursued, given the plethora of articles already published in scientific journals or uploaded to repositories. Various approaches using Bayesian Networks have been employed to determine different aspects induced by this disease, such as diagnosing pneumonia based on images and on clinical and epidemiological data of the patients, among others. The prognosis of such pneumonias, based on clinical data, is also a substantial area of inquiry. However, these aspects will be addressed in subsequent work.

The effectiveness of BNs to identify causal genetic biomarkers of Alzheimer's disease with acceptable precision, evaluated the learned of BNs against the expert opinions of clinicians and entropy calculations, and they developed an ontology-based diagnostic classification, demonstrating the feasibility of direct, automated construction of a high-quality health topology and ontology based on medical records [54]

Formalised a BN-based breast disease diagnosis system; the system retrained itself on new data, resulting in a network benefit, improving the analysis of results [72]models involved an elicitation of experts in the field and a learning phase using an available data set, which makes the decision model more robust and reliable. The BN has the ability to deal with partial observations and uncertainty, which makes the model suitable for the clinical context. Diagnostic clinical decision support system prototype showed some promise, with high levels of validity and reliability; however, issues related to the underlying Bayesian belief network, small sample size, and use of radiological imaging as a gold standard measure were highlighted as requiring further investigation before considering clinical evidence.

With this results BNs demonstrated as a machine learning technique can accurately some diagnose without human intervention, even in mild disease. Incorporation of this tool into electronic medical record systems may aid in the recognition of this underdiagnosed disorders. The results reported by Cruz-Ramírez et al.[49] show a certain element of subjectivity implicitly contained in the data. The findings obtained suggest that observers see different things when looking at the samples under the microscope, a situation that significantly decreases the performance of the Bayesian classifiers used for diagnosing breast cancer. Meanwhile, Djebbar et al.'s [52] experimental results showed that the application of BNs and case-based reasoning improve the efficiency of case retrieval applied to the diagnosis of liver disease.

A web-based medical decision support system optimized to maximize the NPV, thus reducing the cases of undetected risk situations in patients admitted to the emergency room with nonspecific chest pain who are at risk of a heart attack. The model achieved a high NPV for the validation data set. Many aspects of the most mechanism have not been well studied, it is important that the computer program be able to integrate the information scattered across various publications with a variety of different experiments. BNs are particularly attractive for this task due to the flexibility of using Bayesian graphical models for knowledge synthesis [17]. For this application, the quantification of the uncertainty is as important as the prediction itself. A very large standard deviation suggests deficiencies in current knowledge and points to the need for further experiments with the corresponding metabolites. Accordingly, BNs are also advantageous, since they automatically provide information about the range of possible levels of metabolites that are not directly measured.

We showed that the basic advantage of using BN classifiers for predicting diseases, compared to traditional clinical prediction methods based on simple parameters with low prognostic precision, is that they simultaneously use a number of factors associated with symptoms [57]. Among the disadvantages, BN classifiers assume that the attributes are independent, which is not valid in the case of asthma because interactions between some symptoms or patient characteristics could trigger an exacerbation. The advantage of this approach is that it takes into account the dependence that may exist between the attributes. Cai et al. [58] highlight the necessity of obtaining sufficient data in patients to achieve a high predictive accuracy with BNs and suggest that important measures, such as intraoperative blood loss, tumor size, portal vein tumor thrombosis (PVTT), can be applied in medicine to analyze the influence of different variables on a given outcome.

BNs can make probabilistic inferences with any number of observed variables; this property allows us to make predictions with forecast BNs even with incomplete information. As more information becomes available, the forecast can be updated. In the case of limited patient data, the estimated probabilities tend to the global mean of the patient population, while the estimates become more patient-specific as more information is included in the model [59]. Therefore, the Bayesian methodology that underlies BNs encompasses a dynamic notion of forecast, using the probability update based on this new information. Prognostic BNs provide clinicians later in the process with predictions that match the course of earlier phases, for example, a complicated surgical procedure. In addition to adjusted risk estimates, the change in estimated probabilities at earlier forecast times, for instance quantified in terms of risk indices, contains important information about risk progress. Wang et al. [63] described the advantages of using the proposed BN over other reference models: (1) clinicians easily understand the proposed model, (2) it is efficient in both linear and nonlinear modeling, (3) the proposed BN can solve stochastic medical decision-making problems, such as the occurrence of lung cancer brain metastases, by finding the probability distribution of children given the values of parameters in their parents, and vice versa, (4) the model is able to handle situations where information is missing, as indicated in a variety of medical consultations/decisions given partial conditions, and (5) the sensitivity of the proposed Bayesian network is the highest among all the reference methods, and this is essential for identifying cancer patients.

One notable merit of a Bayesian network is that it allows updating probability estimates with new clinical data. Their results suggest that the clinician can intuitively understand the layered correlation between predictors of Bayesian network models. Based on the models, the decision support tool successfully provided probability estimates of post-tibial anterior muscle strength for treating foot drop attributable to lumbar degenerative diseases [65]. These models proved to be robust in internal validation but should be externally validated in other populations. Kaewprag et al.'s [68] BN model offers overall performance comparable to the best classical machine learning algorithms, nearly tripling the sensitivity with only a small cost to specificity and without sacrificing high overall accuracy. This result is considered promising, as high sensitivity may better facilitate preventive care in patients likely to develop a pressure ulcer, which is less expensive than treatment. From a qualitative point of view, strong relationships were identified between risk factors widely recognized as associated with pressure ulcers. Accurate identification of pressure ulcer risk factors is key to understanding the burden of disease and improving care. Clinical collaborators found the Bayesian model useful for identifying dependencies between pressure ulcers and risk factors based on their own experience.

The primary limitations of this review arise from the rapid advancements in Machine Learning and Artificial Intelligence techniques. While some of the referenced articles might appear to be somewhat dated, it is crucial to highlight that Bayesian Networks continue to progress methodologically. Nonetheless, in terms of addressing challenges and framing results, a decade-old publication retains its validity. It is plausible that employing contemporary hybrid networks could yield enhanced accuracy in predictions.

**CONCLUSIONS**

This systematic review of 60 articles indicates that Bayesian networks (BNs) perform adequately, with high levels of accuracy, in the processes of diagnosing and prognosing diseases. Even though advancements in new Machine Learning techniques may yield slight improvements in accuracy levels, the advantage of modeling the process through DAGs significantly aids in better understanding the results and enables the study of causality, something other classification methods are incapable of doing. The growing number of new publications describing their development suggests that prediction models may be becoming increasingly valuable in clinical procedures (according to PubMed, there are 54 articles with the descriptors defined in our search, published in 2023, and more than 90 in WOS). One purpose is to help clinical decision-making by accounting for all patients' specific characteristics to calculate the probability of a certain disorder or problem (diagnosis and prognosis). The review provides evidence supporting the use of BNs over other methods for improving disease prediction.

BNs can be successfully applied to model complex medical problems that require reasoning under conditions of uncertainty. The volume of literature proposing medical BNs and the research effort put into the development of BNs to support medical diagnosis reflects the immense interest in these methods.

Since a BN is versatile, the same model can be used to evaluate, predict, diagnose, and optimize decisions, contributing to the effectiveness of the BN construction effort and the quality of the software on the market [9] In research, a medical diagnostic aid system is implemented to solve decision-making problems with complete data, with the aim of providing doctors with a computer reasoning tool to improve their decisions. The system acts as an expert support system, thanks to the fact that BNs present a knowledge representation methodology to express relationships between variables, using probability theory to handle the problem of medical uncertainty and make a decision understandable to someone other than a specialist.

In terms of future pathways, the development of enhanced preprocessing methods and alternative learning strategies is essential. Furthermore, contemplating the establishment of additional publicly accessible databases should be prioritized in order to train better Bayesian Networks.

Given the considerations outlined above and the exponential growth that Artificial Intelligence has undergone in the past 2-3 years, we assert that persistent research in the field of Bayesian Networks is indispensable (like other machine learning techniques, which yield equal or better results in the field of disease diagnosis and prognosis). This assertion stems from the notable precision obtained in the developed intelligent systems, which aids healthcare professionals in making more expedited decisions with diminished uncertainty. This holds true whether the decisions are based on images or clinical data, spanning a broad spectrum of pathologies.

## ACKNOLEGEMENTS

X.B. would like to thank the Spanish Ministerio de Ciencia e Innovación—Agencia Estatal de Investigación for grant PID2022-136455NB-I00 (jointly financed by the European Regional Development Fund, FEDER).

## AUTHOR CONTRIBUTIONS

Formal analysis, X.B., C.S.; Funding acquisition, D.O. and X.B.; Investigation, X.B., C.S., D.O; JA.Q.; Methodology, X.B., C.S.; Writing—original draft, C.S and X. B; Writing—review & editing, X.B., D.O and JA.Q. All authors have read and agreed to the published version of the manuscript

## CONFLICT OF INTEREST

None declared.

## MULTIMEDIA APPENDIX

We include a supplementary material with the searching strategy.

.